\documentclass[%
 nofootinbib,
 aip,
 jmp,%
 amsmath,amssymb,
preprint,
]{revtex4-1}

\usepackage{graphicx}
\usepackage{dcolumn}
\usepackage{bm}
\usepackage{color}

\begin{document}


\title[Stable interactions in higher derivative field theories of derived type]{Stable interactions in higher derivative field theories of derived type}
\author{V.~A.~Abakumova}
\email{abakumova@phys.tsu.ru}
\affiliation{Physics Faculty, Tomsk State University, Tomsk 634050,
Russia}
\author{D.~S.~Kaparulin}
\email{dsc@phys.tsu.ru}
\affiliation{Physics Faculty, Tomsk State University, Tomsk 634050,
Russia}
\affiliation{Lebedev Institute of Physics, Leninsky ave. 53, Moscow 119991,
Russia}
\author{S.~L.~Lyakhovich}
\email{sll@phys.tsu.ru}
\affiliation{Physics Faculty, Tomsk State University, Tomsk 634050,
Russia}

\date{\today}

\begin{abstract}
We consider the general higher derivative field theories
of derived type. At free level, the  wave operator of derived-type
theory is a polynomial of the order $n\geq 2$ of another operator
$W$ which is of the lower order. Every symmetry of $W$ gives rise to
the series of independent higher order symmetries of the field
equations of derived system. In its turn, these symmetries give rise
to the series of independent conserved quantities. In particular,
the translation invariance of operator $W$ results in the series of
conserved tensors of the derived theory. The series involves $n$
independent conserved tensors including canonical energy-momentum.
Even if the canonical energy is unbounded, the other conserved
tensors in the series can be bounded, that will make the dynamics
stable. The general procedure is worked out to switch on the
interactions such that the stability persists beyond the free level.
The stable interaction vertices  are inevitably non-Lagrangian. The
stable theory, however, can admit consistent quantization. The
general construction is exemplified by the order $N$ extension of
Chern-Simons coupled to  the Pais-Uhlenbeck-type higher derivative
complex scalar field.
\end{abstract}

\maketitle

\section{Introduction}
The higher derivative field theories are notorious for the stability
problems at interacting level, and also in quantum theory, see
\cite{PU, Pavsic, Woodard, Smilga} for discussions and further
references. The most frequently discussed higher derivative field
theories are the modified models of gravity where the stability is
an issue \cite{Salvio, Tomboulis, Belenchia}. Among the modified
gravity theories, $f(R)$ models (for review and further references
see \cite{Sotiriou, Felice}) provide the best known example of
stable nonlinear higher derivative field theory. Some other models
with similar properties can be found in the references
\cite{Smilga1, Bergshoeff2, Nitta}. The stability of this
exceptional class of higher derivative theories is related to the
fact that the canonical energy is bounded because of strong second
class constraints. For discussion of stability in various non-linear
higher derivative mechanical models we refer to \cite{Andrzejewski,
Smilga2, Chen, Pavsic1, Avendao-Camacho} and references therein.

In this paper, we consider a certain class of higher derivative
models which we call \textit{derived theories}. At free level, the
field equations of the derived theory are defined by the higher
derivative wave operator $M$ being a polynomial of another
differential operator $W$. The latter is supposed to be of the first
or second order. As we demonstrate, this class of systems admits,
under certain conditions, inclusion of stable interactions, and the
stability persists at quantum level.

Many well-known higher derivative models fall into the class of
derived theories. For example, the higher derivative scalar field of
the Pais-Uhlenbeck type \cite{PU} is a derived system, where $W$ is
the d'Alembert operator.  Podolsky electrodynamics \cite{Pod} is a
derived system, with the wave operator being a second-order
polynomial of Maxwell operator.  The extended Chern-Simons
\cite{Deser} is a derived theory of the vector field in $3d$
Minkowski space, with the wave operator being a third-order
polynomial in the Chern-Simons operator $*d$, where $*$ is the Hodge
star operator and $d$ is the de Rham differential. The Podolsky and
Chern-Simons electrodynamics have been discussed for many years from
various viewpoints, see \cite{Buffalo1, Buffalo3, Nogueira} and
references therein. In the conformal gravities in 4 and 6 dimensions
\cite{Pope1, Pope2} the linearized equations of motion for spin 2
fields belong to the derived type. The stability  is studied in
these works for the small fluctuations in the vicinity of constant
curvature backgrounds further extending earlier observation of the
work \cite{Maldacena}. With a special choice of boundary conditions,
it is observed that the theory might be stable. In the present work,
we follow a different idea being unrelated to the special boundary
conditions.

The simplest derived-type field equations correspond to the wave
operator being the second-order polynomial in another operator. This
case has been studied in \cite{KLS14}. In this class of models, it
turns out that at free level one can connect two different conserved
quantities to the time-shift symmetry. One of these quantities is
the canonical energy, while another one is a different independent
integral of motion. If the second quantity is bounded, the theory
will be stable at classical level, and the stability can be promoted
to quantum level \cite{KLS14}. In the paper \cite{KKL}, a more
general setup of the derived systems is studied when the wave
operator is a polynomial of arbitrary finite order $n$ in another
operator (see relation (\ref{derived}) in the present work). Once
the primary operator $W$ admits some symmetry, it can be connected
to $n$ conserved quantities of the derived system. In particular, if
one of the symmetries of $W$ is the homogeneity of time, then one
gets $n$ independent conserved quantities, which includes the
canonical energy. In this paper, we adopt slightly more general
setup for the derived systems aimed at relaxing restrictions on
inclusion of stable interactions. The fields are divided into
several subsets such that the wave operator in each subset is a
polynomial in a certain operator. Unlike a more simple case studied
in \cite{KKL}, the primary operators can be different for different
subsets of fields, and the wave operators are defined by different
polynomials for different fields. This more general setup at free
level provides more flexibility for inclusion of stable
interactions.

The stability of interactions in various particular types of higher
derivative theories of derived type has been previously studied in
\cite{KLS14, KKL, AKL18, KKL18}. The proper deformation method has
been suggested in \cite{KLS16} to systematically include stable
interactions in the derived theory which is stable at free level.
The crucial ingredient of the proper deformation method is the
Lagrange anchor. The Lagrange anchor has been first introduced in
the work \cite{KazLS} to BRST embed and quantize not necessarily
Lagrangian dynamics. Later, it has been found \cite{KLS10} that the
Lagrange anchor connects conserved quantities to symmetries for any
system of field equations, be they Lagrangian or not. For the
Lagrangian system the unit operator serves as the Lagrange anchor
that establishes one-tow-one correspondence between symmetries and
conserved currents. In principle, the Lagrange anchor is not
necessarily unique for given system of field equations. Once the
field equations admit multiple Lagrange anchors, the same symmetry
can be connected to different conserved quantities. As is noticed in
the paper \cite{KLS14}, even the simplest derived system, with the
wave operator $M$ being the second-order polynomial of another
differential operator $W$, admits two different Lagrange anchors.
This explains the existence of one more conserved quantity connected
to time-independence besides the canonical energy. Given the
Lagrange anchor, the general method of \cite{KLS16} allows one to
consistently include the interactions into field equations of motion
deforming the conserved quantities connected with the symmetries by
the anchor. If the anchor connects the symmetry with the bounded
quantity, the system remains stable upon inclusion of interaction by
the proper deformation scheme of \cite{KLS16}.

In this paper, we provide a more simple scheme for inclusion stable
interactions in a wider class of derived theories, skipping to
explicitly employ the Lagrange anchor. The basic idea is that we
have two subsets of fields such that the wave operator in each
subset is a polynomial in certain primary operator. The primary
operators are assumed to be Poincar\'e invariant. One of the primary
wave operators is supposed to be gauge invariant. We also assume
that the primary theories admit appropriate covariant interaction
vertex consistent with the gauge symmetry. To construct the stable
interactions between two derived theories, we identify the series of
conserved tensors at free level such that every item is connected to
the space-time translation symmetry. Some of the conserved tensors
can have bounded $00$-component, while the other ones are unbounded.
Then, we seek for the interactions in the derived theories such that
generalize couplings between primary models and keep the appropriate
bounded quantity conserved at interacting level. If the theory
admits bounded conserved quantity, the stability will persist at
interacting level once it is constructed by this method. The
proposed construction can reproduce all the perviously known stable
interactions in higher derivative systems of derived type
\cite{KLS14,KKL,AKL18} and we also add a new example in this paper,
to illustrate the method.

As the illustration of the general method described in the paper, we
construct the consistent and stable interaction between the order
$N$ extension of Chern-Simons and order $2n$ Pais-Uhlenbeck-type
higher derivative complex scalar field. These models have been
discussed earlier at free level in the papers \cite{KLS14, KKL,
AKL18, KKL18} for some specific orders $n$ and $N$. In this paper,
the $N+n$-parameter series of conserved second-rank tensors is
constructed for the free model being connected to the translation
invariance. Also $n$-parameter series of conserved currents is
identified to be connected to a single $U(1)$-symmetry of the
scalar. There are bounded quantities among the conserved observables
that make the theory stable at free level, while the canonical
energy is unbounded. Non-minimal gauge invariant couplings are
identified such that the higher derivative theory remains stable at
interacting level. The stable interaction vertices are inevitably
non-Lagrangian. That does not necessarily obstruct quantization as
we explain in the conclusion.

The article is organized as follows. In the next section, we define
the free derived theories and elaborate on the series of symmetries
and conserved quantities connected to each symmetry of the primary
wave operator $W$. In Section 3, we identify the gauge invariant
couplings such that provide conservation of any fixed representative
of the series of conserved quantities of the derived-type free
system. In Section 4, we consider interactions between the higher
derivative complex scalar field and higher derivative extension of
Chern-Simons. Following general pattern of the previous section, we
explicitly construct the conserved quantities and keep track of the
stability making use of the bounded integrals of motion. In
conclusion, we discuss the results.

\section{Derived-type theories, higher symmetries, and conservation laws}
Consider a set of fields $\phi^a$ in $d$-dimensional Minkowski space
with local coordinates $x^\mu,\mu=0,\ldots,d-1$. We systematically
use the DeWitt condensed notation \cite{DeWitt} in this paper. The
fields are labeled by the condensed indices $a,b, \ldots$. Every
condensed index accommodates all the vector, tensor, spinor,
isotopic, etc. indices and the space-time coordinates. Summation in
the condensed indices implies integration in $x^\mu$. In this
notation, the linear differential operators are represented by
matrices with condensed indices. We assume that the theory admits a
constant metric which is used to raise and lower multi-indices. In
this way, every linear operator has the quadric form. These
quadratic forms correspond to the local functionals which are
bilinear in the fields. We also imply that the fields vanish at the
infinity, so if the quadratic form vanishes in the condensed
notation, this means that the corresponding local functional is the
integral of total divergence.

In the condensed notation, any system of linear field equations
reads:
\begin{equation}\label{linear}
    M_{ab}\phi^b=0\,,
\end{equation}
where $M_{ab}$ is the integral kernel of matrix differential
operator. For the sake of simplicity, we assume that the matrix
$M_{ab}$ is square, so that the number of equations in each
space-time point coincides with the number of fields.  In this class
of theories, $M_{ab}$ is usually called the \emph{wave operator}.
The formal adjoint of the wave operator is defined by
\begin{equation}\label{}
    M^\dagger{}_{ab}=M_{ba}\,.
\end{equation}
The field equations are variational whenever $M^\dagger=M$, in which
case the action functional $S[\phi]$ reads
\begin{equation}\label{L-action}
    S[\phi]=\frac{1}{2}\langle\phi, M\phi\rangle\,,
\end{equation}
where the brackets $\langle\,\,,\,\rangle$ denote the natural pairing between
the fields,
\begin{equation}\label{}
    \langle\phi, M\phi\rangle\equiv\phi^a(M\phi)_a\,,\qquad (M\phi)_a\equiv M_{ab}\phi^b\,,
\end{equation}
and summation  is implied over the repeated multi-indices $a$ when
they stand at different levels. If $M^\dagger_\ast=-M_\ast$ for some
other operator $M_\ast$, its diagonal elements are total
divergencies,
\begin{equation}\label{div-j}
    \int d^dx \partial_\mu j^\mu(\phi)=\langle\phi,
    M_\ast\phi\rangle\,,\qquad \forall\phi\,.
\end{equation}
Once the expression $M_\ast\phi$ vanishes on-shell (\ref{linear}),
the latter formula establishes the relation between conserved
currents and anti-self-adjoint operators.

We say that the variational theory (\ref{linear}) is of the
\textit{derived}-type if the wave operator is a finite-order
polynomial of another self-adjoint operator $W{}^a{}_b$, i.e.
\begin{eqnarray}\label{derived}M(\alpha;W)=\sum_{p=0}^{n}\alpha_{p}W^p\,,\qquad W^\dagger=W\,,\qquad p=2,\ldots,n\,,
\end{eqnarray}
where all the multi-indices are raised and lowered by the metric,
and $\alpha_p\,,\,p=0,\ldots,n\,,$ are some real constants. The
order of the polynomial is assumed to be irreducible, so the
coefficient at the highest order is nonzero, $\alpha_n\neq0$. In
accordance with the definition, each derived theory is defined by
two ingredients: the self-adjoint operator $W{}^a{}_b$ and
finite-order polynomial,
\begin{equation}\label{Charpol}
    M(\alpha;z)=\sum_{p=0}^n\alpha_pz^p\,,
\end{equation}
with $z$ being formal complex-valued variable. We call $M(\alpha;z)$
the \emph{characteristic polynomial}, while $W{}^a{}_b$ is referred
to as the \emph{primary wave operator}. Being considered in itself,
the primary operator defines the \emph{primary free field theory},
\begin{equation}\label{primary}
    W_{ab}\phi^b=0\,.
\end{equation}
As $W_{ab}$ is self-adjoint, the primary theory is variational. In
our article, we mostly deal with the class of theories, where the
primary wave operator does not involve higher derivatives. In this
setting, the higher derivative derived model (\ref{linear}),
(\ref{derived}) can be considered as originating from the lower
order primary theory (\ref{primary}).

In this paper, we consider the linear operator $X{}^a{}_b$ as a
symmetry\footnote{The notion of symmetry can be understood in
various ways. In this section, we provide the simple non-rigorous
understanding, which is sufficient for the purposes of this work.}
of linear equations (\ref{linear}) if it is interchangeable with the
wave operator of theory in the following sense:
\begin{eqnarray}\label{Sym}
    [X,M]{}^a{}_b=Y^a{}_cM{}^c{}_b\,,\qquad [X,M]{}^a{}_b=X{}^a{}_cM{}^c{}_b-M{}^a{}_c X{}^c{}_b\,,
\end{eqnarray}
with $Y{}^a{}_c$ being some other operator. Once this relation is
valid, the operator $X{}^a{}_b$ defines the linear transformation of
the fields such that leaves the mass shell (\ref{linear}) intact,
\begin{eqnarray}\label{}
    \delta_\xi
    (M\phi)^a=\xi(X+Y){}^a{}_b (M\phi)^b\approx0\,,\qquad \delta_\xi\phi^a=\xi X{}^a{}_b\phi^b\,,
\end{eqnarray}
where $\xi$ is the transformation parameter, being some constant.
The sign $\approx$ means equality modulo equations (\ref{linear}).
Each theory admits the trivial symmetries that read
\begin{equation}\label{}
    X{}^{a}{}_b=\widetilde{X}^a{}_cM{}^c{}_b\,,\qquad
    Y{}^a{}_b=[\widetilde{X},M]{}^a{}_b\,,
\end{equation}
where $\widetilde{X}^a{}_b$ can be any \textbf{}operator. The
corresponding transformations of fields vanish on-shell and do not
contain any valuable information about the dynamics of model. We
systematically ignore the trivial symmetries, considering any
symmetry modulo trivial one.

The non-trivial symmetries of linear system are known to form an
associative algebra. The multiplication operation is just a
composition of operators. It is easy to verify the latter fact: for
each pair of symmetries $(X_1){}^a{}_b,(X_2){}^a{}_b$, we get
\begin{eqnarray}\label{}
    &&[X_1X_2,M]{}^a{}_b=(X_1){}^a{}_c[X_2,M]{}^c{}_b+
    [X_1,M]{}^a{}_c(X_2){}^c{}_b\notag\\
    &&=((Y_1){}^a{}_d(X_2)^d{}_c+(X_1){}^a{}_d(Y_2){}^d{}_c+(Y_2)^a{}_c)
    M{}^{c}{}_b\,.
\end{eqnarray}
The elements of associative algebra that commute with the wave
operator form a subalgebra. In this paper, we mostly consider this
subalgebra as it is connected to conserved quantities.

Now, consider the derived equations (\ref{linear}), (\ref{derived}).
Assume that the primary model (\ref{primary}) has the symmetry
$X{}^a{}_b$ such that commutes with the primary wave operator,
\begin{equation}\label{W-Sym}
    [X,W]=0\,.
\end{equation}
Even if the primary theory admits a single symmetry, the associative
algebra of symmetries of derived theory has the two natural
generators -- $X$ and $W$ -- that commute with each other,
\begin{equation}\label{}
    [X,M]=[W,M]=0\,.
\end{equation}
 By composing the primary symmetry $X$ with the degree of the primary wave operator $W$
 we get the symmetry of the derived equations (\ref{derived}):
\begin{eqnarray}\label{Xp-Sym}
   [X_p,M]=0\,,\qquad (X_p){}^a{}_b=X{}^a{}_c(W^p){}^c{}_b\,,\quad p=0,\ldots,n-1\,.
\end{eqnarray}
Only the terms with $p=0,\ldots,n-1$ are relevant, because the
higher powers of primary operator can be absorbed by the wave
operator (\ref{derived}). In this way, for $p\geq n$, the symmetry
reduces on-shell to the symmetry with $p<n$. The generators $X_p$
can be assembled into the $n$-parameter series of \emph{derived
symmetries} of derived model,
\begin{equation}\label{XSer-Sym}
    X{}^a{}_b(\beta)=\sum_{p=0}^{n-1}\beta_p(X_p){}^a{}_b\,,
\end{equation}
with $\beta_p\,,\,p=0,\ldots,n-1\,,$ being real numbers. As is seen,
all the representatives of the series originate from one and the
same symmetry $X$ of primary model.

The symmetry (\ref{Sym}) preserves the action functional
(\ref{L-action}) if its operator is anti-self-adjoint and commutes
with the wave operator,
\begin{equation}\label{}
    [X,M]=0\,,\qquad X^\dagger=-X\,.
\end{equation}
The corresponding conserved current $j^\mu(\phi)$, being quadratic
in the fields, is defined by the condition
\begin{equation}\label{j-char}
    \int d^dx \partial_\mu j^\mu(\phi)=\langle\phi,XM\phi\rangle\,.
\end{equation}
This formula represents the Noetherian relationship between
symmetries and conservation laws. In the class of derived theories,
a single symmetry (\ref{W-Sym}) of primary model (\ref{primary})
determines the series of derived symmetries (\ref{XSer-Sym}). The
associated $n$-parameter series of conserved currents reads
\begin{eqnarray}\label{jSer}
    j^{\mu}(\beta)=\sum_{p=0}^{n-1}\beta_pj_p{}^\mu(\phi)\,,\qquad \int d^dx \partial_\mu
j_p{}^\mu(\phi)=\langle\phi,X_pM\phi\rangle\,,
\end{eqnarray}
where the quantities $j_p{}^\mu, p=0,\ldots,n-1$\,, come from
symmetries (\ref{Xp-Sym}). In particular, $j_0{}^\mu$ represents the
Noether conserved current for the symmetry (\ref{W-Sym}) of primary
model, while $j_p{}^\mu,p=1,\ldots,n-1\,,$ are other quantities.

The simplest possible symmetry of the free field theory is the
translation invariance. The translation generators $\partial_\mu$
are automatically anti-self-adjoint and they commute with the
primary wave operator (\ref{primary}),
\begin{equation}\label{}
    [\partial^\mu,W]=0\,.
\end{equation}
Once the primary wave operator is translation invariant, the derived
theory enjoy $n$-parameter series (\ref{XSer-Sym}) of derived
symmetries originating from this invariance:
\begin{eqnarray}\label{}
    X{}^{\mu a}{}_{b}(\beta)=\sum_{p=0}^{n-1}\beta_p(X_p){}^{\mu a}{}_{b}\,,\qquad (X_p){}^{\mu a}{}_{b}=(\partial^\mu W^p){}^a{}_b\,.
\end{eqnarray}
Each of these symmetries preserves the action (\ref{L-action}) of
the derived model (\ref{linear}), (\ref{derived}). The corresponding
conserved currents (\ref{jSer}) constitute the series of second-rank
energy-momentum tensors
\begin{eqnarray}\label{tSer}
    \mathit\Theta^{\mu\nu}(\beta)=\sum_{p=0}^{n-1}\beta_p
    T_p{}^{\mu\nu}(\phi)\,,\qquad \partial_\nu T_p{}^{\mu\nu}(\phi)=\langle \phi, \partial^\mu W^p
    M\phi\rangle\,.
\end{eqnarray}
This series includes the canonical energy-momentum as
$T_0{}^{\mu\nu}(\phi)$, while $T_p{}^{\mu\nu}, p=1,\ldots,n-1\,,$
are different conserved tensors associated with space-time
translation invariance of the model. In this way, the translation
invariance of the derived-type field theory results in the series of
the conserved tensors.

As the stability of higher derivative model is concerned, the
$00$-component of the conserved tensor is of interest. The
$00$-component of tensor (\ref{tSer}) reads
\begin{equation}\label{t00-Ser}
    \mathit\Theta^{00}(\beta)=\sum_{p=0}^{n-1}\beta_p
    T_p{}^{00}(\phi)\,.
\end{equation}
This expression is given by the sum of canonical energy $T_0{}^{00}$
and the other contributions $T_p{}^{00},p=1,\ldots,n-1$. Even if the
canonical energy is unbounded due to higher derivatives, the
quantity (\ref{t00-Ser}) can define bounded conserved charge. The
bounded conserved quantity, if it exists, stabilizes classical
dynamics of derived model (\ref{linear}), (\ref{derived}) at free
level.

The entries of conserved current series (\ref{jSer}) (or conserved
tensor series (\ref{tSer})) are independent in general, even though
it is not a theorem. This fact is supported by the following
observations: (i) the quantities $j_p{}^\mu $ are bilinear forms in
the fields $\phi^a$ and their space-time derivatives, and (ii) the
total number of derivatives involved in $j_p{}^\mu $ increases with
$p$. Once the symmetry $X$ and primary operator $W$ are the matrix
differential operators of orders $n_X$ and $n_W$, the conserved
current $j_p{}^\mu$ involves at most
\begin{equation}\label{}
    (p+n)n_W+n_X-1
\end{equation}
derivatives of fields. The term with the highest number of
derivatives contributes to the conserved quantity, and it is
on-shell nontrivial at least in the case without gauge symmetry and
constraints. Applying this argument to $j_{n-1}{}^\mu$, we conclude
that the highest order term can't come from the linear combination
of other currents with the lower number of derivatives. Thus,
$j_{n-1}{}^\mu$ is an independent conserved quantity. Proceeding
with the same argument to
 $j_{n-2}{}^\mu, j_{n-3}{}^\mu,\ldots, j_0{}^\mu$, we conclude that all these
quantities are not functions of each other. The maximal number of
independent entries in the conserved current series (\ref{jSer})
equals to the order $n$ of characteristic polynomial
(\ref{Charpol}). In the higher derivative scalar field model the
maximal number of independent conserved quantities is connected with
the space-time translation invariance\cite{KLS14}. In gauge
theories, some generators of series (\ref{jSer}) can be on-shell
trivial. For example, the extended Chern-Simons model of order $n$
admits $n-1$ independent conserved tensors (\ref{tSer})\cite{KKL}.
All the above applies to free theories. At the non-linear level, the
procedure of inclusion of interaction is sensitive to the choice of
representative in the conserved tensor series (\ref{jSer}) which is
conserved at the interacting level. This motivates us to work with
the full series of conserved quantities, even if some of its
generators are dependent.

The discussion above does not address the gauge symmetries. Now we
explain how the gauge symmetry is accounted for in the field
theories of derived type. The wave operator may have a null-vector,
\begin{equation}\label{WA-null}
    M{}^a{}_bR{}^b{}_\alpha=0\,.
\end{equation}
In this case, the derived theory  (\ref{linear}), (\ref{derived}) admits the gauge
transformation such that preserves the mass shell,
\begin{equation}\label{gs-primary}
    \delta_\varepsilon(M{}_{ab}\phi^b)=0\,,\qquad\delta_\varepsilon\phi^a=R{}^a{}_\alpha\varepsilon^\alpha,
\end{equation}
where $\varepsilon^{\alpha}$ are the gauge parameters, being
arbitrary functions of space-time coordinates. The operator
$R{}^a{}_\alpha$ is the gauge symmetry generator. In the present
work, we consider the class of derived models whose gauge symmetries
come from null-vectors of primary model, i.e.
\begin{equation}\label{}
     M{}^a{}_bR{}^b{}_\alpha=0\qquad\Leftrightarrow\qquad W{}^a{}_bR{}^b{}_\alpha=0\,.
\end{equation}
This property is not automatically satisfied for general derived
theory, so this is an additional assumption. The Podolsky's
electrodynamics and the extended Chern-Simons are consistent with
this assumption, for example.

\section{Consistent interactions of derived models}
In this section, we construct the class of consistent interactions
between two Poincar\'e invariant derived theories, with one of them
being gauge. At free level, these theories admit series of conserved
energy-momentum tensors. We construct the interaction such that
provides conservation of the deformation of certain representative
of energy-momentum tensor series. In so doing, we admit not
necessarily Lagrangian vertices.\footnote{For general problems of
inclusion of consistent interactions between not necessarily
Lagrangian theories see \cite{KLS13}.} To solve the problem we
proceed from the assumption that the two primary theories admit
consistent interaction vertex. Then, we seek for the way to lift
this vertex to the level of derived theory in such a way that does
not spoil the stability of the free derived theory.

Consider two subsets of fields $\phi^a\,,\mathit{\Phi}^i$ on
$d$-dimensional Minkowski space, with the multi-indices $a,i$
labeling the fields. The primary operators are denoted by $W_{ab}$
and $\mathcal{W}_{ij}$, so the primary field equations read
\begin{equation}\label{2-primary}
    W_{ab}\phi^b=0\,,\qquad \mathcal{W}_{ij}\mathit{\Phi}^j=0\,.
\end{equation}
These equations are variational, with the action functional being
the sum of actions (\ref{L-action}) of the fields $\phi^a$ and
$\mathit{\Phi}^i$. The primary operators are assumed to be invariant
under the space-time translations,
\begin{equation}\label{}
    [\partial^\mu,W]=[\partial^\mu,\mathcal{W}]=0\,.
\end{equation}
The Poincar\'e symmetry implies the existence of conserved
energy-momentum tensor
\begin{equation}\label{thetaphiPhi}
    \mathit{\Theta}^{\mu\nu}(\phi,\mathit{\Phi})=T^{\mu\nu}(\phi)+T^{\mu\nu}(\mathit{\Phi})\,,
\end{equation}
where $T^{\mu\nu}(\phi)$ and $T^{\mu\nu}(\mathit{\Phi})$ denote the
contributions of free fields $\phi$ and $\mathit{\Phi}$. The
quantity $\mathit{\Theta}^{\mu\nu}$ is determined by the formula
\begin{equation}\label{}
    \int d^dx\partial_\nu\mathit{\Theta}^{\mu\nu}(\phi,\mathit\Phi)=\langle\phi, \partial^\mu W\phi\rangle+
    \langle\mathit\Phi,\partial^\mu\mathcal{W}\mathit\Phi\rangle\,.
\end{equation}
The theory of the field $\mathit{\Phi}^i$ is assumed to be
gauge invariant. The gauge symmetry generator $R^i{}_\alpha$ of the
field is the null-vector of the primary operator $\mathcal{W}_{ij}$,
\begin{equation}\label{mathcalWR}
    \mathcal{W}_{ij}R^{j}{}_\alpha=0\,,
\end{equation}
see Eq. (\ref{WA-null}). The corresponding gauge transformation reads
\begin{equation}\label{gt-P}
    \delta_\varepsilon\mathit{\Phi}^i=R^i{}_\alpha\varepsilon^\alpha,\qquad \delta_\varepsilon\phi^a=0\,,
\end{equation}
where $\varepsilon^\alpha$ are gauge parameters. We do not allow gauge freedom for the
fields $\phi^a$.

We  assume that there exists a variational interaction vertex
between primary models such that (i) the gauge transformation
(\ref{gt-P}) is preserved at the interacting level and (ii) the
action functional is at most quadratic in $\phi$. The most general
action  such that meets these requirements has the form
\begin{equation}\label{S-primary-int}
    S[\phi,\mathit{\Phi}]=\frac{1}{2}(W_{ab}+\mathit{\Gamma}_{ab}(\beta \mathit{\Phi}))\phi^a\phi^b+\frac{1}{2}\mathcal{W}_{ij}\mathit{\Phi}^i\mathit{\Phi}^j\,,
\end{equation}
where the operator of vertex
$\mathit{\Gamma}_{ab}(\beta\mathit{\Phi})$ is a function of quantity
$\beta\mathit{\Phi}$, and $\beta$ is the coupling constant, which
can be arbitrary real number. We assume that
$\mathit{\Gamma}_{ab}(\beta\mathit{\Phi})$ is a finite-order
polynomial in $\mathit\Phi^i$, i.e.
\begin{equation}\label{}
    \mathit\Gamma_{ab}(\beta
    \mathit\Phi)=\sum_{k=1}^{k_{\text{max}}}\beta^k\mathit{\Gamma}^{(k)}{}_{ab}(\mathit{\Phi})\,,\qquad
    k_{\text{max}}<+\infty\,,
\end{equation}
with $\mathit{\Gamma}{}^{(1)}{}_{ab},
\mathit{\Gamma}^{(2)}{}_{ab},\ldots$ being linear, quadratic etc. in
the field $\mathit{\Phi}$. The expression $\beta^k$ means the $k$-th power of $\beta$. The presence of such decomposition
ensures that the equations of motion of the model are finite-order
polynomials in the coupling constant $\beta$. The Lagrange equations for the action functional
(\ref{S-primary-int}) read
\begin{eqnarray}\label{EL-primary-int}
   \partial_a S\equiv W_{ab}(\beta\mathit\Phi)\phi^b=0\,,\qquad \partial_i S\equiv \mathcal{W}_{ij}\mathit{\Phi}^j+\beta J_i(\phi,\beta\mathit\Phi)=0\,,
\end{eqnarray}
where
\begin{eqnarray}\label{W-inv}
W_{ab}(\beta\mathit\Phi)=W_{ab}+\mathit{\Gamma}_{ab}(\beta\mathit{\Phi})\,,\qquad
J_{i}(\phi,\beta\mathit\Phi)=
\partial_i\mathit\Gamma_{ab}(\beta\mathit\Phi)\phi^a\phi^b\,,
\end{eqnarray}
and the  derivatives $\partial_a, \partial_i$ are understood as
variational in $\phi^a,\Phi^i$. As is seen, in the model with
interaction, the operator of vertex
$\mathit\Gamma_{ab}(\beta\mathit\Phi)$ is added to free equations of
originally non-gauge field $\phi^a$, while the gauge field theory
get additional current-like contribution
$J_i(\phi,\beta\mathit\Phi)$. This structure of non-linear theory is
typical for lower spin fields. For example, it includes the minimal
coupling between the electromagnetic field and the charged matter
fields.

The infinitesimal gauge transformation of the action functional
(\ref{S-primary-int}) reads
\begin{equation}\label{gt-pA}
    \delta_\varepsilon\phi^a=\beta R{}^{a}{}_{b\alpha}\phi^b\varepsilon^\alpha\,,\qquad
    \delta_\varepsilon \mathit\Phi^i=R{}^{i}{}_{\alpha}\varepsilon^\alpha\,,
\end{equation}
where $R{}^i{}_\alpha$ is the null-vector of primary operator $\mathcal{W}_{ij}$
(\ref{mathcalWR}), and $R^a{}_{b\alpha}$ is some field-independent
structure function. We assume that it is antisymmetric in the
indices $ab$,
\begin{equation}\label{R-R}
 R{}_{ab\alpha}=-R{}_{ba\alpha}\,.
\end{equation}
The invariance of action (\ref{S-primary-int}) with respect to the
gauge transformations (\ref{gt-pA}) is equivalent to the condition
\begin{eqnarray}\label{G-W-R}
     (W+\mathit{\Gamma}(\beta\mathit{\Phi}))^{a}{}_cR^c{}_{b\alpha}-R^a{}_{c\alpha}(W+\mathit{\Gamma}(\beta\mathit{\Phi}))^c{}_b+\,R^i{}_\alpha\partial_i\mathit{\Gamma}{}^a{}_b(\beta\mathit{\Phi})=0\,.
\end{eqnarray}

As the gauge symmetries and gauge identities of Lagrangian theories
are connected by the Noether theorem, conditions (\ref{R-R}), (\ref{G-W-R}) determine the
selection rule for the structure functions
$\mathit{\Gamma}{}^{a}{}_{b},R^{a}{}_{b\alpha}$. The meaning of these conditions
is that the wave operator $W_{ab}(\beta\mathit\Phi)$ (\ref{W-inv}) is gauge invariant in the following sense:
\begin{eqnarray}\label{[R,W]}
    [W(\beta\mathit\Phi), R_\alpha]_{ab}&+&R^i{}_\alpha\partial_i\mathit{\Gamma}{}_{ab}(\beta\mathit{\Phi})=0\,,\notag\\\,
    [W(\beta\mathit\Phi), R_\alpha]_{ab}&=&W{}_{ac}(\beta\mathit\Phi)R^{c}{}_{b\alpha}-R_{ac\alpha}W{}^c{}_b(\beta\mathit\Phi)\,,\qquad\forall\beta\,.
\end{eqnarray}
The latter relation immediately ensures the gauge invariance of equations of motion of the field $\phi^a$.

The translation invariance of interacting theory
(\ref{S-primary-int}) is understood in the following sense:
\begin{eqnarray}\label{[p,W]}
   [W(\beta\mathit\Phi),\partial^\mu]{}_{ab}&+&\beta\partial^\mu\mathit\Phi^i\partial{}_i\mathit\Gamma{}_{ab}(\beta\mathit\Phi)=0\,,\notag\\\,
   [W(\beta\mathit\Phi), \partial^\mu]_{ab}&=&W{}_{ab}(\beta\mathit\Phi)\partial^\mu-\partial^\mu W{}_{ab}(\beta\mathit\Phi)\,,\qquad \forall\beta\,.
\end{eqnarray}
Once this condition is met, the action functional
(\ref{S-primary-int}) is preserved by the space-time translations.
The energy-momentum tensor of the non-linear theory has the
following structure:
\begin{equation}\label{ThetaSer}
    \boldsymbol{\mathit{\Theta}}^{\mu\nu}(\phi,\mathit{\Phi})=T^{\mu\nu}(\phi,\mathit\Phi)+T^{\mu\nu}(\mathit{\Phi})\,.
\end{equation}
The expression $T^{\mu\nu}(\phi,\mathit\Phi)$ is the gauge invariant
extension of the tensor $T^{\mu\nu}(\phi)$ (\ref{thetaphiPhi}) of
free theory, while $T^{\mu\nu}(\mathit{\Phi})$ is the
energy-momentum of free gauge field. The defining condition for the
conserved tensor reads
\begin{equation}\label{}
    \int d^d x\partial_\nu\boldsymbol{\mathit{\Theta}}^{\mu\nu}(\phi,\mathit\Phi)=
    \langle\phi, \partial^\mu W(\beta\mathit\Phi)\phi\rangle+
    \langle\mathit\Phi,\partial^\mu\mathcal{W}\mathit\Phi\rangle\,.
\end{equation}
The rhs of this expression is a total divergence because
$\partial^\mu W(\beta\mathit\Phi), \partial^\mu\mathcal{W}$ are
anti-self-adjoint operators.

Let us explain the meaning of conditions (\ref{R-R}), (\ref{G-W-R})
and its impact on the structure of interactions. Assume that the
gauge transformation of the gauge field $\mathit{\Phi}^i$ has a
null-mode $\varepsilon=\overline{\varepsilon}(x)$ such that
\begin{equation}\label{}
    R{}^i{}_\alpha\overline{\varepsilon}^\alpha=0\,.
\end{equation}
In this case, the action functional (\ref{S-primary-int}) is
preserved by the following transformation:
\begin{equation}\label{rigid-sym}
    \delta_{\overline{\varepsilon}}\phi^a=\xi
    R{}^a{}_{b\alpha}\overline{\varepsilon}^\alpha\,,\qquad
    \delta_{\overline\varepsilon}\mathit\Phi^i=0\,,
\end{equation}
with the parameter $\xi$ being a constant. In the free limit, this
transformation corresponds to certain internal symmetry of the free
model of the fields $\phi^a$, which localizes at the interacting
level. This symmetry does not follow from the relativistic
invariance of equations (\ref{2-primary}), so we have additional
prerequisite for constriction of interaction at free level. As we
deal with internal symmetry, this gives a restriction on the
multiplet of fields $\phi^a$ that are involved into the interaction.

The current-like term $J_i$ (\ref{W-inv}) originates from the
internal symmetry (\ref{rigid-sym}). As a consequence of relations
(\ref{R-R}), (\ref{G-W-R}), it is gauge invariant and satisfies
gauge identity,
\begin{equation}\label{R-J}
    \delta_\varepsilon J_i=0\,,\qquad R^i{}_\alpha J_i+R{}^a{}_{b\alpha}\partial_a S\, \phi^b=0\,.
\end{equation}
If the internal symmetry is the $U(1)$-transformation, and gauge
generator $R{}^i{}_\alpha$ is gradient, the current-like term
literally corresponds to the current of the $U(1)$-charge,
\begin{equation}\label{}
    J_i=J_\mu(x)\,,
\end{equation}
which meets condition (\ref{R-J}). Once current-like term $J_i$ is
added to the equations of motion of the gauge field, and the primary
operator $W$ is replaced by its gauge invariant extension
$W(\beta\mathit\Phi)$\,, the non-linear theory remains gauge invariant.
This means that interactions (\ref{EL-primary-int}) imitate the
electromagnetic-like couplings between gauge and matter fields.

Let us now turn to the details of inclusion of interactions between
derived theories. The most general ansatz for two derived theories
with the primary operators $W$ and $\mathcal{W}$ has the form
\begin{equation}\label{2-derived}
    M_{ab}(\alpha;W)\phi^b=0\,,\qquad \mathcal{M}{}_{ij}(A;\mathcal{W})\mathit{\Phi}^j=0\,.
\end{equation}
The characteristic polynomials of derived models (\ref{2-derived})
are the most general of orders $n$ and $N$, respectively,
\begin{equation}\label{2-Charpol}
    M(\alpha;z)=\sum_{p=0}^{n}\alpha_{p}z^p\,,
    \qquad
    \mathcal{M}(A;z)=\sum_{q=1}^{N}A_qz^q\,,
\end{equation}
where $z$ is a formal complex-valued variable. The real numbers
$\alpha_p\,,\, p=0,\ldots,n\,,$ and $A_q\,,\,q=1,\ldots,N\,,$ are model
parameters that distinguish different theories in
the considered class.

The derived theories (\ref{2-derived}) have the same space-time and gauge
symmetries as primary models. In particular, models (\ref{2-derived}), (\ref{2-Charpol}) are invariant under
gauge transformation (\ref{gt-P}). As for space-time translation,
a single symmetry of primary theory induces the $n+N$-parameter
series of derived symmetries such that
\begin{eqnarray}\label{}
    X^{\mu a}{}_b(\beta)=\sum_{p=0}^{n-1}\beta_p(\partial^\mu W^p){}^a{}_{b}\,,\qquad X^{\mu
i}{}_j(B)=\sum_{q=0}^{N-1}B_q(\partial^\mu\mathcal{W}^q){}^i{}_j\,,
\end{eqnarray}
where the real numbers $\beta_p,B_q$ are parameters. The
corresponding set of the conserved quantities is given by the series
of energy-momentum tensors
\begin{equation}\label{Theta1}
    \mathit{\Theta}^{\mu\nu}(\beta,B)=\sum_{p=0}^{n-1}\beta{}_p T{}_p{}^{\mu\nu}(\phi)+\sum_{q=0}^{N-1}B{}_qT{}_q{}^{\mu\nu}(\mathit{\Phi})\,,
\end{equation}
where the quantities $T{}_p{}^{\mu\nu}(\phi)$, $T{}_q{}^{\mu\nu}(\mathit\Phi)$ are
defined by the conditions
\begin{eqnarray}\label{d-Theta-free}
    \int d^d x\partial_\nu\mathit{\Theta}^{\mu\nu}(\beta,B)&=&\sum_{p=0}^{n-1}\beta{}_p \langle\phi,\partial^\mu W^pM\phi\rangle+\sum_{q=0}^{N-1}B{}_q
\mathit\langle\mathit\Phi,\partial^\mu\mathcal{W}^q\mathcal{M}\mathit\Phi\rangle\,.
\end{eqnarray}
By construction, the canonical energy-momentum
tensor of the model is included in the series as
\begin{equation}\label{}
\mathit{\Theta}^{\mu\nu}_{\text{can}}(\phi,\mathit{\Phi})=
T{}_0{}^{\mu\nu}(\phi)+T{}_0{}^{\mu\nu}(\mathit{\Phi})\,.
\end{equation}
Even though the canonical energy-momentum is almost always unbounded
due to higher derivatives, the other bounded quantities can present
in the series. Free higher derivative theory can be stabilized
by these quantities.

By interaction between two gauge models (\ref{2-derived}) we mean a
formal deformation of free theory with coupling parameter $\beta$
such that the non-linear equations of motion read
\begin{eqnarray}\label{E}
    &&E_{a}(\beta)\equiv M_{ab}\phi^b+\sum_{k=1}^\infty\beta^k
    \mathit{\Gamma}{}^{(k)}{}_{a}(\phi,\mathit{\Phi})=0\,,\notag\\
    &&\mathcal{E}_{i}(\beta)\equiv \mathcal{M}_{ij}
    \mathit{\Phi}^j+\sum_{k=1}^{\infty}\beta^k\mathit{\Gamma}^{(k)}{}_{i}(\phi,\mathit{\Phi})=0\,.
\end{eqnarray}
The structure functions $\mathit{\Gamma}{}^{(k)}$ are assumed to
have the homogeneity degree $k+1$ in the variables
$\phi^a,\mathit{\Phi}^i$, so $\mathit{\Gamma}{}^{(1)}$ are quadratic
in the fields, $\mathit{\Gamma}{}^{(2)}$ are cubic, and etc. No
action principle is required for interacting theory, so non-linear
equations (\ref{E}) can be non-Lagragian. We consider interaction
(\ref{E}) consistent if the number of gauge symmetries and gauge
identities is preserved by coupling and at least one representative
of the free series of energy-momentum tensors (\ref{d-Theta-free})
is still conserved at the non-linear level.

The $n+N$-parameter series of interaction vertices for free
equations (\ref{2-derived})  can be presented in the following form
\begin{eqnarray}\label{e-int-general}
    E_a\equiv\overline{M}{}_{ab}(\beta,B)\phi^b=0\,,\qquad\mathcal{E}_i\equiv
\mathcal{M}_{ij}\mathit{\Phi}^j+\sum_{p=0}^{n-1}\beta_p(\overline{J}_{p}){}_{i}(\beta,B)=0\,,
\end{eqnarray}
where the notation is used:
\begin{eqnarray}\label{Wbar}
    \overline{M}_{ab}(\beta,
    B)=\sum_{p=0}^{n}\alpha{}_{p}\overline{W}^p(\beta,B)\,,\qquad
    (\overline{J}_{p}){}_{i}(\beta, B)=\frac{1}{2}\partial_i\langle\phi,\overline{W{}}^p(\beta,B)\overline{M}(\beta,B)\phi\rangle\,,
\end{eqnarray}
and the line above means that the gauge field $\mathit\Phi^i$ is replaced by more general expression:
\begin{eqnarray}\label{Pbar}
    \overline{W}(\beta,B)=\left.W(\beta\mathit\Phi)\right|_{\beta\mathit\Phi=\overline{\mathit\Phi}(\beta,B)}\,,\qquad \overline{\mathit{\Phi}}{}^i(\beta,B)=\beta\mathit\Phi^i+\sum_{q=1}^{N-1}B_{q}(\mathcal{W}^q\mathit{\Phi}){}^i\,.
\end{eqnarray}
The coupling constants are real numbers
$\beta\,,\,\beta_{p}\,,\,B_q\,,\,p=0,\ldots,n-1\,,\,q=1,\ldots,N-1\,$.
Interactions (\ref{e-int-general}) generalize couplings of the
primary theory (\ref{EL-primary-int}) in the following sense: the
free wave operator $M$ of the fields $\phi^a$ is replaced by its
gauge invariant extension $\overline{M}(\beta,B)$, and the series of
current like terms $(\overline{J}_{p}){}_{i}(\beta, B)$ is added to
the free gauge equations.

The following facts ensure consistency of interaction (\ref{e-int-general}):
(i) the equations of motion are preserved by the gauge transformation
 (\ref{gt-pA}), while transformation law for the equations of motion reads
\begin{equation}\label{gt-eom}
    \delta_\varepsilon E^{a}=\beta R^{a}{}_{b\alpha}E^b\varepsilon^\alpha\,,\qquad \delta_\varepsilon \mathcal{E}^i=0\,;
\end{equation}
(ii) there are the gauge identities between equations of motion,
\begin{equation}\label{g-id}
    R{}^i{}_\alpha
    \mathcal{E}_i+\sum_{p=0}^{n-1}\beta_pR_{ab\alpha}(\overline{W}{}^p){}^b{}_cE^a\phi^c\equiv0\,;
\end{equation}
(iii) the second-rank conserved tensor $\boldsymbol{\mathit\Theta}^{\mu\nu}$ is defined by the condition
\begin{eqnarray}\label{d-Theta}
    \int d^dx \partial_\nu\boldsymbol{\mathit\Theta}^{\mu\nu}(\beta,B)=\sum_{p=0}^{n-1}\beta_p
    \langle\phi,\partial^\mu\overline{W}{}^pE\rangle
    +\sum_{q=0}^{N-1}B_q \langle\mathit\Phi, \partial^\mu \mathcal{W}^q \mathcal{E}\rangle\,,\qquad B_0\equiv\beta\,.
\end{eqnarray}
The proof of relations (\ref{gt-eom}), (\ref{g-id})  uses identities
\begin{equation}\label{App-ids}\begin{array}{c}
    (\text{a})\,\delta_\varepsilon (\overline{W}{}^p\phi)^a=\beta R^{a}{}_{b\alpha}(\overline{W}{}^p\phi)^b\varepsilon^\alpha\,;
    \quad(\text{b})\,
    R^i{}_\alpha(\overline{J}_{p}){}_{i}+R_{ab\alpha}(\overline{W}{}^p){}^b{}_cE^a\phi^c\equiv0\,;\\[3mm](\text{c})\,
    \delta_\varepsilon (\overline{J}_{p}){}_{i}=0\,,
\end{array}\end{equation} which hold for $p=0,\ldots,n-1$. We deduce these
relations in the Appendix A. The conserved tensor
$\boldsymbol{\mathit\Theta}^{\mu\nu}$ (\ref{d-Theta}) is a
deformation of a selected representative of free energy-momentum
series (\ref{d-Theta-free}), whose parameters are defined by
coupling constants. In Appendix B, we ensure the rhs of Eq.
(\ref{d-Theta}) is a total divergence and establish the structure of
conserved quantity.

Let us comment on the origin of interactions (\ref{e-int-general}).
The model (\ref{2-derived}) admits $n$-parameter series of gauge
invariant current-like terms
\begin{eqnarray}\label{J-1}
    J_i(\beta)=\sum_{p=0}^{n-1}\beta_p(J_{p}){}_{i}(\beta)\,,\qquad (J_{p}){}_{i}(\beta)=\frac{1}{2}\partial_i\langle\phi,W{}^p(\beta\mathit{\Phi})M(\beta\mathit{\Phi})\phi\rangle\,.
\end{eqnarray}
Once the primary wave operator $W$ is replaced by its gauge
invariant extension $W(\beta\mathit\Phi)$, the inclusion of such
current-like terms preserves gauge invariance. This means that
theory (\ref{e-int-general}) with the gauge-invariant primary
operator (\ref{W-inv}) and current-like term (\ref{J-1}) is
consistent. At this ways, we conclude that there is $n+1$-parameter
series of interactions with coupling constants
$\beta,\,\beta_0,\ldots,\beta_{n-1}\,.$ To involve coupling
constants $B_q$ into account, we perform the shift of the gauge
field $\beta\mathit\Phi\mapsto\overline{\mathit\Phi}(\beta,B)$
(\ref{Pbar}). This step preserves consistency of interactions and
inserts $N-1$ coupling parameters $B_1,\ldots,B_{N-1}$\, into the
field equations.

The non-linear theory (\ref{e-int-general}) is stable if the bounded
representative of free energy-momentum tensor series (\ref{Theta1})
is still conserved at the interacting level. The latter requirement
can be interpreted as the selection rule for admissible couplings.
The derived theories, whose characteristic polynomial
(\ref{Charpol}) has non-degenerate real roots, are usually stable at
free level\footnote{Several known examples confirm this observation
\cite{KKL},\cite{KKL18},\cite{AKL18}, though it is not a theorem at
the moment.}. This gives a good chance for existence of stable
interactions between such models. The theories with multiple real
or/and complex roots of the characteristic polynomial usually have
the degrees of freedom whose dynamics is unbounded by any conserved
quantity. These degrees of freedom are the true ghosts unless they
are suppressed by constraints or gauged out. The stability of
interactions in the models having no bounded conserved quantity at
free level is a subtle issue, which we do not consider in the
present paper.

The general representative in the class of interacting theories
(\ref{e-int-general}) is non-Lagrangian as the equations of motion
are not given by the variational derivatives of any functional. The
field equations are Lagrangian if the values of coupling parameters
meet the condition
\begin{equation}\label{Lag-vertex}
    \beta_0=\beta\,,
\end{equation}
and all other constants vanish. The action functional reads
\begin{equation}\label{}
    S=\frac{1}{2}\Big(\langle \phi,M(\beta\mathit\Phi)\phi\rangle+\langle\mathit{\Phi},\mathcal{M}\mathit{\Phi}\rangle\Big)\,.
\end{equation}
The corresponding conserved quantity is the canonical
energy-momentum tensor of the model. In the class of higher
derivative derived equations, the variational interaction vertex
usually suffers from the Ostrogradski instability. In contrast to
Lagrangian interactions, the non-Lagrangian couplings can preserve
bounded representative in the series of free conserved quantities
(\ref{Theta1}). These couplings define the stable non-linear theory.

Let us summarize the results of this section. Relations
(\ref{e-int-general})-(\ref{Pbar}) provide a general receipt of
construction of series of gauge invariant interaction vertices
between two derived theories,  if the gauge invariant  coupling is
known for primary models. The structure of interaction vertex
resembles couplings between electromagnetic field and charged
matter: the wave operator of originally non-gauge field is replaced
by its gauge invariant extension, while the current-like term is
added to equations of motion of gauge field. If the characteristic
polynomials of derived theories have the orders $n$ and $N$, there
are $n+N$ coupling constants. Any specific choice of coupling
constants selects certain representative in the series of free
conserved energy-momentum such that still conserves at the
interacting level. If the free theory has bounded conserved
quantities, it admits the stable interactions, though the stable
couplings are non-Lagrangian.

\section{Couplings between extended Chern-Simons and higher derivative scalar}
Following the general pattern of the previous section,  let us
consider the interactions between the gauge vector field and the
charged scalar. In this case, we have two subsets of fields on $3d$
Minkowski space,
\begin{equation}
\phi^a=(\text{Re}\,\phi)(x)+i(\text{Im}\,\phi)(x)\,, \qquad
\mathit{\Phi}^i=\mathit{\Phi}^\mu(x)\,.
\end{equation}
The primary operators read
\begin{equation}\label{primopSChS}
W=m^{-2}\partial_\mu\partial^\mu\,, \qquad
\mathcal{W}_{\mu\nu}=m^{-1}\varepsilon_{\mu\rho\nu}\partial^\rho\,.
\end{equation}
To make contact with the general scheme of previous section, we
identify the d'Alembertian as the primary operator for the scalar.
For the vector field, the Chern-Simons operator serves as the
primary operator. The constant $m>0$ has the dimension of mass. Here
we use it to make the primary operators dimensionless. The primary
equations of motion,
\begin{equation}\label{primSChS}
m^{-2}\partial_\mu\partial^\mu\phi=0\,, \qquad
m^{-1}\varepsilon_{\mu\nu\rho}\,\partial^\nu\mathit{\Phi}^\rho=0\,,
\end{equation}
have the gauge symmetry (\ref{gt-P}), with the gauge generator for
the vector field being gradient, $R^i{}_\alpha\equiv\partial^\mu
$\,.

The primary models (\ref{primSChS}) admit obvious variational
interaction vertex,
\begin{equation}\label{SSChS}
\displaystyle S[\phi,\mathit{\Phi}]=\frac{1}{2}\int\big[\phi^*D_\mu
D^\mu\phi+m\varepsilon_{\mu\nu\rho}\mathit{\Phi}^\mu\partial^\nu\mathit{\Phi}^\rho\big]\,d^3x\,,
\end{equation}
where $D_\mu$ is the covariant derivative,
\begin{equation}\label{Dmuphi}
D_\mu(\beta)\phi=\big(\partial_\mu-i\beta\mathit{\Phi}_\mu\big)\phi\,.
\end{equation}
For the complex conjugate field $\phi^*$, the covariant derivative
is given by the complex conjugation of (\ref{Dmuphi}). The action
functional (\ref{SSChS}) has form (\ref{S-primary-int}) with the
operator of vertex,
\begin{eqnarray}
\mathit{\Gamma}&\equiv&D_\mu
D^\mu-\partial_\mu\partial^\mu=(\partial_\mu-i\beta\mathit{\Phi}_\mu\big)(\partial^\mu-i\beta\mathit{\Phi}^\mu\big)-\partial_\mu\partial^\mu\,,
\end{eqnarray}
being the second-order polynomial in the coupling constant
$\beta$\,. The Lagrange equations for the action functional
(\ref{SSChS}) read
\begin{eqnarray}
\frac{\delta S}{\delta\mathit{\Phi}^\mu}\equiv
m\varepsilon_{\mu\nu\rho}\partial^\nu\mathit{\Phi}^\rho+i\beta\big(\phi^*D_\mu\phi-\phi
D_\mu\phi^*\big)=0\,,\qquad \frac{\delta S}{\delta
\phi}\equiv\frac{1}{2}D_\mu D^\mu\phi^*=0\,.
\end{eqnarray}
These equations have form (\ref{EL-primary-int}) with
\begin{eqnarray}
J_i\equiv J_\mu(\phi, \beta\mathit{\Phi}) =i(\phi^*D_\mu\phi-\phi
D_\mu\phi^*\big)\,,\qquad W(\beta\mathit{\Phi})=D_\mu D^\mu\,.
\end{eqnarray}
As the gauge field is a vector, the current-like term holds world
index. As we can see, the quantity $J_\mu(\phi, \beta\mathit{\Phi})$
is the charge current of complex scalar field. The infinitesimal
gauge transformation (\ref{gt-pA}) of the action functional
(\ref{SSChS}) reads
\begin{equation}\label{gtSChS}
\delta\phi=i\beta\phi^*\varepsilon\,, \qquad
\delta\mathit{\Phi}^\mu=\partial^\mu\varepsilon\,.
\end{equation}
The canonical energy-momentum tensor (\ref{ThetaSer}) for the action
(\ref{SSChS}) has the form
\begin{equation}\begin{array}{c}
\displaystyle
\boldsymbol{\mathit{\Theta}}_{\mu\nu}=F_{\mu}F_{\nu}-\frac{1}{2}\eta_{\mu\nu}F_{\lambda}F^{\lambda}+D_\mu\phi^*D_\nu\phi+D_\nu\phi^*D_\mu\phi
-\eta_{\mu\nu}D_\lambda\phi^*D^\lambda\phi\,, \\[3mm] F_\mu\equiv
m\varepsilon_{\mu\nu\rho}\partial^\nu\mathit{\Phi}^\rho\,.
\end{array}\end{equation} It is obvious that its $00$-component is bounded.
So, the theory of interacting Chern-Simons and charged scalar is
stable at the level of primary theories.

Now we consider interactions between two derived theories, whose
primary operators are given by (\ref{primopSChS}).
 The free equations (\ref{2-derived}) are chosen in the form
\begin{eqnarray}\label{derSChS}
&&M(\alpha; W)\,\phi\equiv m^2\prod\limits_{p=0}^{n-1}\big(m^{-2}\partial_\mu\partial^\mu+\alpha_p^2\big)\phi=0\,,\notag\\
&&\mathcal{M}_{\mu\nu}(A;\mathcal{W})\,\mathit{\Phi}^\nu\equiv\frac{m^2}{2}\sum\limits_{q=1}^NA_q(\mathcal{W}^q)_{\mu\nu}\mathit{\Phi}^\nu=0\,.
\end{eqnarray}
Here we assume that all the roots of characteristic polynomial
(\ref{Charpol}) for scalar field are different and positive. We
ignore all the other options because they do not result to the
stable theory at the free or interacting level. The extended
covariant derivative $\overline{D}_\mu$ is defined as follows:
\begin{equation}\label{overlineDmu}
\displaystyle
\overline{D}_\mu(\beta,B)\phi=\big(\partial_\mu-i\sum_{q=0}^{N-1}B_q(\mathcal{W}^q)_{\mu\nu}\mathit{\Phi}^\nu\big)\phi\,,
\qquad B_0\equiv \beta\,.
\end{equation}
The gauge invariant extension of the free wave operator $M(\alpha;
W)$ (\ref{derSChS}) reads
\begin{equation}
\overline{M}(\alpha; \overline{W}(\beta\mathit{\Phi}))\,\phi\equiv
m^2\prod\limits_{p=0}^{n-1}\big(m^{-2}\overline{D}_\mu\overline{D}^\mu+\alpha_p^2\big)\phi=0\,.
\end{equation}
The series of current-like terms (\ref{J-1}) can be represented in
the form
\begin{equation}
\displaystyle
(J_p)_\mu=i\big(\phi^{*(p)}\overline{D}_\mu\phi^{(p)}-\phi^{(p)}\overline{D}_\mu\phi^{*(p)}\big)\,,
\, p=0,\ldots,n-1\,,
\end{equation}
where the notation is used:\footnote{The fields $\phi^{(p)}$
describe irreducible components of reducible representation of the
Poincar\'e group that is described by the higher derivative theory
of complex scalar. The vectors $(J_p){}_{\mu}\,,\,
p=0,\ldots,n-1\,,$ are charge currents of irreducible components,
while the coupling constants $\beta_p,\,p=0,\ldots,n-1\,,$ can be
interpreted as the charges of components.}
\begin{equation}\label{phip}
\displaystyle \phi^{(p)}=\prod_{\substack{p'=0\\ p'\neq
p}}^{n-1}\frac{m^{-2}\overline{D}_\mu
\overline{D}^\mu+\alpha_{p'}^2}{\alpha_p^2-\alpha_{p'}^2}\phi\,,\,\,
p=0,\ldots,n-1\,.
\end{equation}
Let us notice that $(J_p)_\mu$ are the linear combinations of
$(J_{p}){}_{i}$ in (\ref{J-1}). We deal with objects $(J_p)_\mu$ for
reasons of convenience.

The $n+N$-parameter series of interaction vertices
(\ref{e-int-general}) for free equations (\ref{derSChS}) read
\begin{widetext}
\begin{eqnarray}\label{ESChS}
&&\displaystyle E\equiv m^2\prod\limits_{p=0}^{n-1}\big(m^{-2}\overline{D}_\mu(\beta,B)\overline{D}^\mu(\beta,B)+\alpha_p^2\big)\phi=0\,,\notag\\
&&\displaystyle
\mathcal{E}_\mu\equiv\frac{m^2}{2}\sum\limits_{q=1}^NA_q(\mathcal{W}^q)_{\mu\nu}\mathit{\Phi}^\nu+i\sum\limits_{p=0}^{n-1}\beta_p\big(\phi^{*(p)}\overline{D}_\mu(\beta,B)\phi^{(p)}-\phi^{(p)}\overline{D}_\mu(\beta,B)\phi^{*(p)}\big)=0\,,
\end{eqnarray}
\end{widetext}
where $\overline{D}_\mu(\beta,B)$ is the covariant derivative
(\ref{overlineDmu}), $\phi^{(p)}$ is given by (\ref{phip}), and
$\beta\,,\,\beta_p\,,\,B_q\,,\,p=0,\ldots,n-1\,,\,q=1,\ldots,N-1\,,$
are the coupling parameters.

Equations (\ref{ESChS}) are obviously gauge invariant with respect
to gauge symmetry (\ref{gtSChS}). The gauge identity between
equations (\ref{ESChS}) is seen in the form
\begin{equation}
\displaystyle
\partial^\mu\mathcal{E}_\mu-i\sum\limits_{p=0}^{n-1}\prod_{\substack{p'=0\\
p'\neq
p}}^{n-1}\frac{\beta_p}{\alpha_p^2-\alpha_{p'}^2}\big(\phi^{*(p)}E-\phi^{(p)}E^*\big)=0\,.
\end{equation}
The series of the energy-momentum tensors (\ref{Theta1}) for
non-linear theory (\ref{ESChS}) has the structure
\begin{equation}\label{ThetaSChS}
\boldsymbol{\mathit{\Theta}}_{\mu\nu}(\phi,\mathit{\Phi})=\sum_{p=0}^{n-1}\beta_p(T_p){}_{\mu\nu}(\phi,\mathit{\Phi})+\sum_{q=0}^{N-1}B_q(T_q)_{\mu\nu}(\mathit{\Phi})\,,
\end{equation}
where the quantities $(T_p){}_{\mu\nu},\,p=0,\ldots,n-1\,,$ and
$(T_q){}_{\mu\nu},\,q=0,\ldots,N-1\,,$ read
\begin{equation}\label{TqChS}\begin{array}{c}\displaystyle
(T_p){}_{\mu\nu}(\phi,\mathit{\Phi})=\overline{D}_\mu\phi^{*(p)}\overline{D}_\nu\phi^{(p)}+
\overline{D}_\nu\phi^{*(p)}\overline{D}_\mu\phi^{(p)}-\eta_{\mu\nu}\overline{D}_\lambda\phi^{*(p)}\overline{D}^\lambda\phi^{(p)}+
m^2\alpha_p^2\eta_{\mu\nu}\phi^{*(p)}\phi^{(p)}\,,\\[3mm]\displaystyle
(T_q){}_{\mu\nu}(\mathit{\Phi})=\sum_{r,s=1}^{N-1}\sum_{q'=1}^{N}C_{r,s}^{q,q'}A_{q'}\big(\mathit{\Phi}^{(r)}{}_{\mu}\mathit{\Phi}^{(s)}{}_{\nu}+\mathit{\Phi}^{(r)}{}_{\nu}\mathit{\Phi}^{(s)}{}_{\mu}-\eta_{\mu\nu}\mathit{\Phi}^{(r)}{}_\lambda\mathit{\Phi}^{(s)\lambda}\big)\,,
\\[3mm]\displaystyle
\mathit{\Phi}^{(q)}{}_{\mu}\equiv(\mathcal{W}^q){}_{\mu\nu}\mathit{\Phi}^{\nu}\,.
\end{array}\end{equation} The constants $C_{r,s}^{q,q'}$ are defined as
follows:
\begin{equation}\label{Crsqq}
C_{r,s}^{q,q'}= \left\{
\begin{array}{rll}
-1, &  q>q'\,, &q>r,s\,,\\
&  &q+q'=r+s+1\,;\\
1, & q<q'\,, &q\leq r,s\,,\\
& &q+q'=r+s+1\,;\\
0, & \text{otherwise}\,.&\\
\end{array}
\right.
\end{equation}
As is seen from (\ref{ThetaSChS}), the  quantity
$\boldsymbol{\mathit\Theta}^{\mu\nu}$ is the deformation of selected
representative of energy-momentum tensor series of free complex
scalar and Chern-Simons,
\begin{eqnarray}\label{ThetaSChS-free}
\boldsymbol{\mathit{\Theta}}^{\text{free}}_{\mu\nu}(\phi,\mathit{\Phi})&=&
\sum_{q=0}^{N-1}B_q(T_q)_{\mu\nu}(\mathit{\Phi})+
\sum_{p=0}^{n-1}\beta_p(\partial_\mu\phi^{*(p)}\partial_\nu\phi^{(p)}\notag\\
&+&\partial_\nu\phi^{*(p)}\partial_\mu\phi^{(p)}
-\eta_{\mu\nu}\partial_\lambda\phi^{*(p)}\partial^\lambda\phi^{(p)}+m^2\alpha_p^2\eta_{\mu\nu}\phi^{*(p)}\phi^{(p)})\,.
\end{eqnarray}
In this formula all the parameters of the conserved quantity are
fixed by the interaction, so a single representative of free series
of conserved quantities survives at the interacting level.

Let us discuss the stability of the theory (\ref{ESChS}) at the
interacting level. The $00$-component of (\ref{ThetaSChS}) can be
written in the form
\begin{eqnarray}\label{Theta00SChS}
\displaystyle \boldsymbol{\mathit{\Theta}}_{00}(\phi,\mathit{\Phi})&=&\frac{m^2}{2}\sum\limits_{r,s=1}^{N-1}C_{r,s}(A,B)\mathit{\Phi}^{(r)}{}_{\mu}\mathit{\Phi}^{(s)}{}_{\mu}\notag\\
&+&\sum\limits_{p=0}^{n-1}\beta_p\big(\overline{D}_\mu\phi^{*(p)}\overline{D}_\mu\phi^{(p)}+m^2\alpha_p^2\phi^{*(p)}\phi^{(p)}\big)\,.
\end{eqnarray}
where the matrix $C_{r,s}(A,B)$ is defined by
\begin{equation}\label{C-matrix}
\displaystyle
C_{r,s}(A,B)=\sum_{q=0}^{N-1}\sum_{q'=1}^{N}C_{r,s}^{q,q'}B_qA_{q'}\,.
\end{equation}
The expression (\ref{Theta00SChS}) is a quadratic form in the
variables $\mathit{\Phi}^{(q)}{}_{\mu}\,, \phi^{(p)}$\,. It is
bounded if $C_{r,s}(A,B)\,,\, r,\,s=1,\ldots,N-1\,,$ is a positive
definite matrix, and $\beta_p>0\,,\,p=0,\ldots,n-1\,,$ are positive
numbers.\footnote{The quantity $C_{r,s}(A,B))$ is the Bezout matrix
of the characteristic polynomial (\ref{Charpol}) of free extended
Chern-Simons theory and characteristic polynomial of derived
symmetry (\ref{Xp-Sym}), (\ref{XSer-Sym}). It is defined by the
relation
\begin{equation*}
\sum_{r,s=1}^{N-1}C_{r,s}(A,B)z^{r}u^{s}=\frac{M(A;z)N(B;u)-M(A;u)N(B;z)}{z-u}\,,
\end{equation*}
where
\begin{equation*}
 M(A;z)=\sum_{q=1}^NA_qz^q\,, \qquad
N(B;z)=\sum_{q=0}^{N-1}B_{q+1}z^{q+1}\,.
\end{equation*}
Accounting this definition, formula (\ref{Crsqq}) takes the form
\begin{equation*}
C_{r,s}^{q,q'}=\frac{\partial^2 C(A,B)}{\partial A_q
\partial B_{q'}}\,.
\end{equation*}
}

We consider the stability condition of interacting theory as the
natural selection rule for admissible couplings. In the sector of
scalar field, it restricts the coupling parameters $\beta_p$. In the
sector of gauge field, the matrix $C_{r,s}(A,B)$ can be positive
definite or indefinite depending on the values of the free model
parameters $A_q$ and coupling constants $B_q$. In the work
\cite{KKL} the following fact is noticed about the free extended
Chern-Simons theory: the series of conserved quantities
(\ref{TqChS}) includes the bounded representative if the
characteristic polynomial of the theory (\ref{derSChS}) has simple
real nonzero roots and a zero root of multiplicity one or two. In
all the other cases, the theory is unstable. For the viewpoint of
representation theory, the stability condition requires from the
free extended Chern-Simons theory to describe the unitary
representation of the Poincar\'e group. The roots of characteristic
equation determine the masses of irreducible representations in this
case. Once the roots of the characteristic equation of  extended
Chern-Simons theory are such that any conserved quantity is
unbounded in the series (\ref{TqChS}), and therefore the stable
interactions are impossible to include, the free theory corresponds
to the non-unitary representation. It seems quite natural that the
non-unitary theory cannot be stable at interacting level.

In general, the interaction (\ref{ESChS}) is non-Lagrangian. The
Lagrangian interaction vertex corresponds to the following values of
coupling parameters:
\begin{equation}
\beta=\beta_0\,, \,\,
B_1=\ldots=B_{N}=\beta_1=\ldots=\beta_{n-1}=0\,.
\end{equation}
The conserved quantity for variational couplings is the canonical
energy-momentum tensor.  As the canonical energy
$(\boldsymbol{\mathit{\Theta}}_\text{can}\,)_{00}$ is unbounded,
the Lagrangian interaction is unstable. The field equations
(\ref{ESChS}) include stable interactions, so the corresponding
coupling is non-Lagrangian. The coupling parameters that lead to
stable interactions are determined by conditions
$\beta_p>0\,,p=0,\ldots,n-1\,,$ and positive definiteness requirement for
the form $C_{r,s}(A,B)$ (\ref{C-matrix}).

\section{Conclusion and discussion of results}
At first, let us summarize the basic assumptions about their
consequences for the class of higher derivative field theories we
study in this paper. We consider inclusion of stable interactions in
certain class of higher derivative field theories which we term as
\textit{derived} models. The derived higher derivative theory
implies that the same field admits free field equation with the wave
operator $W$ without higher derivatives. We call this model a
\textit{primary} theory. The wave operator of the derived theory is
a $n$-th order polynomial in $W$. We call it a characteristic
polynomial of the derived theory. Every symmetry of primary theory
results in the $n$-parametric series of symmetries of the derived
theory. These symmetries are connected to the series of independent
conserved quantities. In particular, the translation invariance of
the primary model results in the $n$-parametric series of conserved
tensors. Under certain assumptions about the roots of characteristic
polynomial, the series can include bounded conserved quantities. The
canonical energy-momentum is included into the series, though it is
always unbounded. Once the free derived theory admits bounded
conserved quantities, it is considered stable.

We consider inclusion of interactions between two different derived
field theories. One of these is supposed to be gauge invariant, and
another one is non-gauge. We assume that the primary theories of
these two models admit consistent and stable interactions, with the
gauge symmetry remaining abelian at interacting level. We also
impose certain technical restrictions on the interaction between
primary theories as explained in the section 3. In particular, these
restrictions exclude the "gauging" scenarios of inclusion of
interactions\footnote{The "gauging" schemes mean that the gauge
symmetry of self-interacting gauge fields can change upon inclusion
of the non-gauge matter fields. Concerning the literature on
gauging, we can mention\cite{gaug1, gaug2} and references therein.
}. In principle, these restrictions could be relaxed without
breaking the general scheme of inclusion of stable interactions
in the derived-type theories, though this would make the
consideration much more cumbersome. Once the primary theories admit
consistent interactions, these can be lifted to the level of derived
theory. The lift is not unique, we get $n+N$-parametric series of
consistent interactions, where $n,N$ are the orders of
characteristic polynomials of the two models. At free level, the two
derived theories admitted $n+N$ parametric series of the conserved
tensors. Upon inclusion of interaction, only one tensor conserves
with any fixed set of the interaction parameters. The canonical
energy momentum is conserved if the interaction is Lagrangian. As
the canonical energy is unbounded, the Lagrangian interaction is
unstable.  Once the free derived theory admits bounded conserved
quantity, its conservation can be always preserved with appropriate
consistent and Poincar\'e invariant interaction, so the stability
persists at interacting level, though with the non-Lagrangian
vertex.

Also notice that even if the interaction is included in such a way
that none of the bounded conserved quantities (from the series
identified in section 2)  of the free theory remains conserving at
interacting level, these non-conserved currents can be still
informative. So they can be still relevant for Lagrangian
interactions. If the dynamics is known of bounded non-conserved
quantity, than we learn about the compact evolving phase-space
surface where the dynamics is confined. This might be useful for
identifying the isles of stability, and evaluating the velocity of
running away solutions. For discussion of these ideas and further
references see the recent work \cite{Nicolas}.

Even though the vertices of stable interactions are always
non-Lagrangian in this class of higher derivative theories, they can
still admit quantization. The matter is that these non-Lagrangian
field equations can admit Lagrange anchor which allows one to
quantize the non-Lagrangian field theory. The examples are provided
in the paper \cite{KLS14} where the some special interactions are
considered between the derived model with the quadratic
characteristic polynomial, and the theory without higher
derivatives. Another side of the existence of the Lagrange anchor is
that the theory should admit Hamiltonian formalism even though the
higher derivative equations are non-Lagrangian. The explicit
examples of Hamiltonian formalism for third-order non-Lagrangian
derived systems can be found in papers \cite{AKL18, KKL18}. The
constrained Hamiltonian formalism for $n$-th order derived theory
with the stable interactions constructed in Section 4 is developed
in the recent work \cite{AKL18-RFJ}.

\begin{acknowledgments} We thank
P.~Kazinski, A.~Kamenshchik, A.~Sharapov, and A.~Vikman for
discussions. The work of VAA is supported by Tomsk State University
Competitiveness Improvement Program. The work of DSC is supported by
the Russian Science Foundation grant 18-72-10123 in association with
the Lebedev Physical Institute of RAS. The work of SLL is supported
by the project 3.5204.2017/6.7 of Russian Ministry of Science and
Education.
\end{acknowledgments}

\appendix

\section{The proof of identities (\ref{App-ids})}

Let us mention at first the obvious fact we make use of below. The
anti-commutator $\{\, ,\, \}$ of self-adjoint and anti-self-adjoint
operators is anti-self-adjoint. This means, the diagonal elements
elements vanish of the anti-commutator
\begin{equation}\label{anti-id}
    \langle\phi\,,\{\overline{W}{}^p\,,R_\alpha\}\phi\rangle\equiv0\,,\qquad p=0,1,2\ldots\,.
\end{equation}
If the space-time translation, being  the anti-self-adjoint
operator, is substituted in this relation, the diagonal element will
vanish. This means, we get the integral of a total divergence
\begin{equation}\label{d-sig}
    \langle\phi\,,\{\overline{W}{}^p\,,\partial^\mu\}\phi\rangle=\int d^dx\partial_{\nu}\mathit\Sigma_{p}{}^{\mu\nu}(\phi,\mathit\Phi)\,,\, p=0,1,2\ldots\,,
\end{equation}
with $\mathit\Sigma_{p}{}^{\mu\nu}$ being some second-rank tensor.

The relation (\ref{App-ids}.a) is proved by induction. The statement
is obviously true for $p=0$. Assuming that (\ref{App-ids}.a) holds
for some unspecified value of $p=k$ and using (\ref{G-W-R}), we find
\begin{eqnarray}\label{}
    \delta_\varepsilon (\overline{W}{}^{k+1}\phi){}^a&=&\beta\varepsilon^\alpha (\overline{W}{}^{a}{}_{c}R^{c}{}_{b\alpha}+
    R{}^i{}_\alpha\partial_i\mathit\Gamma{}^a{}_b)(\overline{W}{}^{k}\phi){}^b=\beta\varepsilon^\alpha
R^a{}_{b\alpha}(\overline{W}{}^{k+1}\phi)^b\,.
\end{eqnarray}
As $k$ is arbitrary positive number, the statement holds for any nonnegative $p=0,1,2,\ldots$\,.

Consider relation (\ref{App-ids}.b). Using the Leibnitz rule to compute variational derivative,
we can represent the quantity $(\overline{J}_{p}){}_{i}$ in the following form
\begin{equation}\label{}
    (\overline{J}_{p}){}_{i}=\frac{1}{2}\sum_{k=0}^{n}\sum_{l=1}^{p+k}\alpha_k\langle\phi,\overline{W}^{l-1}\overline{\partial_{i}\mathit\Gamma}{}\,\overline{W}^{p+k-l}\phi\rangle\,.
\end{equation}
Contracting this relation with the gauge generator $R{}^i{}_\alpha$, we get
\begin{equation}\label{RJ}
    R{}^i{}_\alpha(\overline{J}_{p}){}_{i}=\frac{1}{2}\sum_{k=0}^{n}\sum_{l=1}^{p+k}\alpha_k\langle\phi,\overline{W}^{l-1}
    R{}^i{}_\alpha\overline{\partial_{i}\mathit\Gamma}{}\,\overline{W}^{p+k-l}\phi\rangle\,.
\end{equation}
With the account of relations (\ref{anti-id}) and equations of
motion (\ref{e-int-general}), we see that
\begin{eqnarray}\label{RE}
    R_{ab\alpha}(\overline{W}{}^p){}^b{}_c\phi^cE^a=\langle\phi,\overline{M}\overline{W}R{}_{\alpha}\phi\rangle=\frac{1}{2}\sum_{k=0}^n\alpha_k\langle\phi,[\overline{W}{}^{p+k},R_\alpha]\phi\rangle\,.
\end{eqnarray}
The latter quantity can be expressed as a sum of commutators,
\begin{equation}\label{}
    [R_\alpha, \overline{W}^{p+k}]=\sum_{l=1}^{p+k}\overline{W}^{l-1}[R_\alpha, \overline{W}]\overline{W}{}^{p+k-l}\,.
\end{equation}
Substituting this expression into (\ref{RE}) and adding the result to (\ref{RJ}), we obtain
\begin{eqnarray}\label{}
    R{}^i{}_\alpha (\overline{J}_{p}){}_{i}+R_{ab\alpha}(\overline{W}{}^p){}^b{}_c\phi^cE^a=\frac{1}{2}\sum_{k=0}^{n}\sum_{l=1}^{p+k}
    \alpha_k\Big\langle\phi,\overline{W}^{l-1}\Big([\,\overline{W},R_\alpha]+R{}^i\partial_i\mathit\Gamma\Big)\overline{W}^{p+k-l}\phi\Big\rangle\,.
\end{eqnarray}
Due to identity (\ref{[R,W]}), this expression vanishes identically. This proves the formula (\ref{App-ids}.b).

Now consider the issue of gauge invariance of current-like term
$(\overline{J}_{p}){}_{i}$. Computing the gauge variation of the expression (\ref{RJ}) and
using formula (\ref{App-ids}.a), we obtain
\begin{eqnarray}\label{}
\delta_\varepsilon
(\overline{J}_{p}){}_{i}&=&\frac{1}{2}\beta\varepsilon^\alpha\sum_{k=0}^{n}\sum_{l=1}^{p+k}\alpha_k\Big\langle\phi,\overline{W}{}^{l-1}
    \partial_{i}\Big([\overline{W},R_\alpha]+R{}^j{}_\alpha\partial_{j}\mathit\Gamma\Big)\overline{W}{}^{p+k-l}\phi\Big\rangle\,.
\end{eqnarray}
Again, this expression vanishes identically as a result of identity
(\ref{[R,W]}). This proves formula (\ref{App-ids}.c).

\section{Existence and structure of second-rank conserved tensor $\boldsymbol{\mathit\Theta}^{\mu\nu}(\beta,B)$}

In this appendix, we prove that the rhs of equation (\ref{d-Theta})
is the conserved tensor. Substituting the equations of motion
(\ref{e-int-general}) into this relation, using the definition of
current-like term (\ref{Wbar}), we get

\begin{equation}\label{74}\begin{array}{l}\displaystyle
    \int
    d^dx\partial_{\nu}\boldsymbol{\mathit\Theta}^{\mu\nu}(\beta,B)=\\[3mm]\displaystyle
    =\sum_{k=0}^{n}\sum_{p=0}^{n-1}\alpha_k\beta_p
    \Big\langle\phi,\Big(\partial^\mu\overline{W}{}^{p+k}+\,\frac{1}{2}\overline{\mathit\Phi}{}^i\partial_i\partial^\mu\overline{W}{}^{p+k}\Big)\phi\Big\rangle
    +\sum_{q=0}^{N-1}B{}_q\langle\mathit\Phi{},\partial^\mu\mathcal{M}\mathit\Phi\rangle\,.
\end{array}\end{equation} The second term defines the series of
energy-momentum tensors of free theory of gauge field
(\ref{d-Theta-free}),
\begin{eqnarray}\label{}
   \partial_\nu
   T^{\mu\nu}(\mathit\Phi)=\sum_{q=0}^{N-1}B{}_q\mathit\Phi{}^{i}(\partial^\mu\mathcal{M}\mathit\Phi)_i\,,\qquad
   T^{\mu\nu}(\mathit\Phi)=\sum_{q=0}^{N-1}B_qT_q{}^{\mu\nu}(\mathit\Phi)\,.
\end{eqnarray}
It remains to prove that the first term is a divergence of some tensor. Integrating by parts, we decompose the entries of the sum into
three contributions
\begin{equation}\label{}\begin{array}{c} \displaystyle
\frac{1}{2}\Big\{-\Big\langle\phi,\Big([\overline{W}{}^{p+k},\partial^\mu]
+\partial^\mu\overline{\mathit\Phi}{}^i\partial_i\overline{W}{}^{p+k}\Big)\phi\Big\rangle+\Big\langle\phi,\{\overline{W}{}^{p+k},\partial^\mu\}\phi\Big\rangle+\\[3mm]
\displaystyle+\int d^d
x\partial^\mu\Big\langle\phi,\overline{\mathit\Phi}{}^i\partial_i\overline{W}{}^{p+k}\phi\Big\rangle
\Big\}\,, \end{array}\end{equation} where the square brackets
$[\,,\,]$ and braces $\{\,,\,\}$ denote the commutator and
anti-commutator of enclosed operators. The first term vanishes
identically due to (\ref{[p,W]}). The second term is the total
divergence (\ref{d-sig}), as well as the third one. The consistency
of formula (\ref{d-Theta}) is proven.

The conserved tensor $\boldsymbol{\mathit\Theta}^{\mu\nu}(\beta,B)$ has the following structure:
\begin{eqnarray}\label{}
    &&\boldsymbol{\mathit{\Theta}}{}^{\mu\nu}(\beta,B)=\sum_{k=0}^{n}\sum_{p=0}^{n-1}
    \alpha_k\beta_p\Big(\mathit\Sigma_{p+k}{}^{\mu\nu}(\phi,\mathit\Phi)\notag\\
    &&+\,\frac{1}{2}\eta^{\mu\nu}\Big\langle\phi,\overline{\mathit\Phi}{}^i\partial_i\overline{W}{}^{p+k}\phi\Big\rangle\Big)+
    \sum_{q=0}^{N-1}B_qT_q{}^{\mu\nu}(\mathit\Phi)\,,
\end{eqnarray}
where $T_q{}^{\mu\nu}(\mathit\Phi)$ are the energy-momentum tensors of free gauge field, and $\mathit\Sigma_{p+k}{}^{\mu\nu}(\phi,\mathit\Phi)$ is
defined in (\ref{d-sig}). As in the free limit
\begin{eqnarray}\label{}
T_p{}^{\mu\nu}(\phi)=\sum_{k=0}^{n}
\alpha_k\mathit\Sigma_{p+k}{}^{\mu\nu}\,,\qquad\Big\langle\phi,\overline{\mathit\Phi}{}^i\partial_i\overline{W}{}^{p+k}
\phi\Big\rangle=0\,,
\end{eqnarray}
the quantity $\boldsymbol{\mathit\Theta}^{\mu\nu}(\beta,B)$ is a
deformation of an exclusive representative in the series of
energy-momentum  tensors (\ref{d-Theta-free}).

\end{document}